# CRITICAL UTILITY INFRASTRUCTURAL RESILIENCE

Giovanna Dondossola[1], Geert Deconinck[2], Felicita Di Giandomenico[3], Susanna Donatelli[4], Mohamed Kaâniche[5], Paulo Verissimo[6]

CESI[1], KUL[2], CNR-ISTI[3], CNIT[4], LAAS-CNRS[5], FCUL[6]

**Keywords:** infrastructure interdependencies, power control system protection, system analysis and model development, resilient control architectures

## Abstract

The paper refers to CRUTIAL[7], *CRitical UTility InfrastructurAL Resilience,* a European project within the research area of Critical Information Infrastructure Protection, with a specific focus on the infrastructures operated by *power utilities*, widely recognized as fundamental to national and international economy, security and quality of life. Such infrastructures faced with the recent market deregulations and the multiple interdependencies with other infrastructures are becoming more and more vulnerable to various threats, including accidental failures and deliberate sabotage and malicious attacks.

The subject of CRUTIAL research are small scale networked ICT systems used to control and manage the electric power grid, in which artifacts controlling the physical process of electricity transportation need to be connected with corporate and societal applications performing management and maintenance functionality. The peculiarity of such ICT-supported systems is that they are related to the power system dynamics and its emergency conditions.

Specific effort need to be devoted by the Electric Power community and by the Information Technology community to influence the technological progress in order to allow commercial intelligent electronic devices to be effectively deployed for the protection of citizens against cyber threats to electric power management and control systems. A well-founded know-how needs to be built inside the industrial power sector to allow all the involved stakeholders to achieve their service objectives without compromising the resilience properties of the logical and physical assets that support the electric power provision.

[1] CESI RICERCA S.p.A. Networks and Infrastructures Department
Via Rubattino 54, Milan. giovanna.dondossola@cesiricerca.it
[2] Electa Katholieke Universiteit Leuven Department of Electrical Engineering
Kasteelpark Arenberg 10, Leuven. Geert.deconinck@esat.kuleuven.be
[3] Consiglio Nazionale delle Ricerche Institute of Science and Information Technologies
Via G. Moruzzi 1, Pisa. felicita.digiandomenico@isti.cnr.it
[4] Consorzio Nazionale Interuniversitario per le Telecomunicazioni Torino Research Unit
C.so Svizzera 185, Torino. susi@di.unito.it
[5] LAAS-CNRS, Université de Toulouse 7 Avenue du Colonel Roche, Toulouse. mohamed.kaaniche@laas.fr
[6] Faculty of Sciences University of Lisboa Department of Informatics
Campo Grande, ED. C6, Piso 3, Lisboa. pjv@di.fc.ul.pt

# Introduction

The problem of security and dependability, or generically speaking, *resilience* [1] of Internet-oriented infrastructure systems, such as web server compounds, is reasonably well understood. Although it is not completely mastered (for example, denial of service is still a research subject), it is receiving adequate attention. However, such is not the case with the problem of resilience of *critical utility infrastructures*. This problem is not completely understood, mainly due to the hybrid composition of these infrastructures.

The process control of utility infrastructures is based on the SCADA (Supervisory Control and Data Acquisition) systems which yield the operational ability to acquire data, supervise and control whatever is the business in question (electricity, water, gas, telecomm). However, they also have interconnections to the standard corporate intranets, and hence indirectly to the Internet (e.g., remote access via dedicated or public networks). The aforementioned SCADA systems were classically not designed to be widely distributed and remotely accessed, let alone be open. They grew-up standalone, closed, not having security in mind.

Widely distributed monitoring, protection and control systems, implementing special protection scheme over the power transmission network, are emerging whose architecture is based on open communication infrastructures.

This opening that we observe nowadays is an afterthought in the line of the generic trend of any informatics system. Whilst it seems non-controversial that such a status quo brings a certain level of threat, namely but not only through interference, we know of no work that has tried to equate the problem by defining a model of "modern utilities distributed systems architecture". We believe that evaluation work on such a model will let us learn about activity patterns of interdependencies that will reveal the potential for far more damaging fault/failure scenarios than those that have been anticipated up to now. Moreover, such a model will be highly constructive as well, for it will form a structured framework for: conceiving the right balance between prevention and removal of vulnerabilities and attacks, and tolerance of remaining potential intrusions and designed-in faults.

In fact, this hap hazardous evolution led to the inevitable: access to operational networks such as remote SCADA maneuvering, ended up entangled with access to corporate intranets and public Internet, without there being computational and resilience models that understand (*represent*) this situation and deal with the resulting interference. In consequence, unlike what exists in simpler, more homogeneous settings, e.g. classical web-based server infrastructures on Internet, it is in most circumstances not possible to devise a dependability and security case for these interconnected critical utility infrastructures.

The damage perspectives that may result from this exposure are overwhelming. They range from wrong maneuvering, to malicious actions coming from terminals located outside, somewhere in the Internet. The targets of these actions are computer control units, embedded components and systems, that is, devices connected to operational hardware (e.g., water pumps and filters, electrical power generators and power protections, dam gates, etc.). In the electrical power provision these situations have already been experimented by citizens in various part of the world. As a single example, among blackouts that occurred in the summer of 2003 in several countries, we can remember the North American one, which is very relevant to explain the project motivations: as highlighted in the analysis report [2], it was the failure of various information systems that thwarted the utility workers' ability to contain the blackout before it cascaded out of control. A type of failure that is characteristic of interdependent critical infrastructures happened then, that is an escalating failure[8], a type of failure that we shall address in this project.

---

[8] "An escalating failure occurs when an existing disruption in one infrastructure exacerbates an independent disruption of a second infrastructure, generally in the form of increasing the severity or the time for recovery or restoration of the second failure".

The main challenge of the project is to make power control resilient in spite of threats to their information and communication infrastructures. Considering the crucial role of control systems in governing the quality and the stability of the electric power service, it is considered of great importance for the utilities operating the infrastructures to dispose of tools for analyzing threat impacts and of technologies for avoiding, or limiting, most serious consequences.

The paper presents the basic research challenges that the CRUTIAL consortium has identified, an overview on the state of the art about the major topics involved, and the work plan that CRUTIAL intends to develop during the project course, based on the modeling and architectural approaches which represent the background of the project team.

## Research challenges and objectives
The project focuses on the electrical power infrastructure and the information infrastructures, by considering different topology realms and different kinds of risks: distinguishing the backbone from the specific information networks and from the infrastructures dedicated to the control and monitoring of the electric power infrastructure, as they usually have different levels of protection; distinguishing faults of different kinds and severities, such as electric power outages and cyber attacks.

The main objective of the project is the investigation of models and architectures that cope with the scenario of openness, heterogeneity and evolvability endured by electrical utilities infrastructures, in the present and near future. The approach taken should support the analysis and management of interdependencies and of the resulting overall operational risk.

Firstly, the project aims to develop comprehensive modeling approaches, supported by measurement based experiments, to analyze critical scenarios in which internal or external faults in a segment of the information infrastructure provoke a serious impact on the controlled electric power infrastructure, in order to understand and master such interdependencies to avoid escalating and cascading failures that result in outages and blackouts. Effort will be focused on the *modeling and analysis* of interdependencies, especially the various types of failures that can occur in the presence of accidental and malicious faults affecting the information and electric power infrastructures: physical (hardware) faults of various temporal characteristics from transient to permanent; yet undiscovered design faults in software ("bugs") and in hardware ("errata"); interaction faults due to mistakes of human operators or maintainers; malicious human-caused faults, i.e. denial or unauthorized alteration of services by software attacks or physical sabotage.
Given the complexity of the analysis task, a difficulty will be to find the right abstractions of the models. The aim is therefore to produce, from conceptual analysis, generic models that can be refined and instantiated. The abstractions will be substantiated by examples taken from the electric application domain. Another expected difficulty is the heterogeneity of the models, given the very different nature of the various components under study.

Secondly, the project will investigate distributed architectures dedicated to the control and management of the power grid, in the perspective of improving the capability of analyzing critical scenarios and designing dependable interconnected power control systems. The architectures under study will address requirements coming from the needs of flexible electric power services, characterized by dispersed energy resources, on-demand control and inputs from the market.
In consequence, our objective is to devise *new architectural configurations* that address the increase in operational risk derived from the analysis made above. This risk derives not only from accidental faults or wrong maneuvers, but also, and very importantly, from both the degree of vulnerability and the level of threat to which the infrastructures and services are subjected. The objective of preventing escalating failures on the various information infrastructures (monitoring, control, management) that interact on a decentralized power grid can only be met by the combined use of fault prevention and tolerance, and by the simultaneous addressing of accidental and malicious faults, also called intrusion-tolerance, enhanced by the provision of on-line monitoring support to evaluate possible alternative architectural configurations in uncertain and evolving scenarios.

**Background**
This section gives an overview of international initiatives and working groups related to the topic of our project, as well as related work on dependability and security modeling and on resilient distributed real-time architectures.

International initiatives and working groups
There is a consensus in the literature on critical infrastructures that interdependency analyses and models constitute a necessary step [3]. The International CIIP Handbook 2004 [4] is a comprehensive collection of information about the various initiatives undertaken by the different countries on the theme of Critical Information Infrastructure Protection (CIIP), mainly at governmental level. The CIIP Handbook underlies the need of developing methodologies for analyzing interdependencies and guiding the protection of critical information infrastructures. In Italy, the *Working Group on Critical Information Infrastructure Protection* released a document [5] in October 2003 describing many elements of the Italian infrastructures, emphasizing their interdependencies, suggesting CIIP policy strategies and research challenges.
In the Unites States many research initiatives and activities related to the protection of critical infrastructures have been undertaken from the nineties. NERC has organized a Cyber Security Urgent Action, resulting in some guidelines, compliance audits, and activities such as workshops for awareness raising. In 1998 the USA's Department of Energy assigned to NERC the role of reference point for cyber security for the electric power sector. The CIPC (Critical Infrastructure Protection Committee) was created to develop and maintain capabilities to respond to security threats and incidents, and support the production of standards and guidelines. Also in the US, the Electric Power Research Institute (EPRI) started the "Infrastructure Security Initiative" addressing power system security at both electrical and cyber levels. The Department of Energy (DOE) published the "21 Steps to Improve Cyber Security of SCADA Networks", whilst the Sandia National Laboratories developed a research program on SCADA electronic security [6].
In Europe the need of setting up research programs on critical infrastructures has been recognized from several years. However a limited number of European Projects has addressed the emerging problems of digitalized critical infrastructures. The ACIP (Analysis & Assessment for Critical Infrastructure Protection) project produced a roadmap about R&D activities to be performed on the topic. A roadmap on research activity in the area of information system dependability has been produced by the AMSD accompanying measure (A Dependability Roadmap for the Information Society in Europe). The FP5-IST research project SAFEGUARD (Intelligent Agents Organization to Enhance Dependability and Survivability of Large Complex Critical Infrastructure) proposed an agent-based architecture for supporting the supervision and decision support systems in critical infrastructural domains.

From the standardization side, there are several working groups devoted to the application of Information Technology in Electric Power Systems. A paper [7] providing a survey on ongoing activities has been published by the Cigré Joint Working Group D2/B3/C2-01 "Security for Information Systems and Intranets in Electric Power Systems".

The American Gas Association (AGA) are active in the development of cryptographic standards for SCADA communication [8].
The Instrumentation, Systems and Automation Society (ISA) published two reference reports [9] on Electronic Security in Manufacturing and Control Systems Environment.

The Computer Security Resource Center, National Institute of Standards and Technology (NIST), including the Process Control Security Requirements Forum (PCSRF), published relevant papers on the topic ([10], [11], [12]).

We can observe that any document by such standard committees emphasizes the importance of Cyber Risk Assessment of power control systems and a first methodological approach has been presented in a recent paper of the Cigrè Group [13].

Also a number of institutional/research organization are performing activities which provide information suitable to the identification of robust architectural patterns for industrial control systems.

The National Infrastructure Security Co-ordination Centre – NISCC in London coordinates a forum on SCADA and Control Systems Information Exchange. They also are going to establish the CERT-UK for SCADA and they are developing the SCADA Incident Database.

The British Columbia Institute of Technology – BCIT is developing a platform for testing SCADA vulnerability. They perform vulnerability analysis of SCADA protocols, they investigate on the use of firewall in SCADA systems and are developing the Industrial Security Incident Database.

The Idaho National Engineering and Environmental Laboratory - INEEL/SNL are developing the SCADA National Test Bed, which is specifically dedicated to testing control systems of a wide area electric power grid.

A test bed for the simulation of attack scenarios to power control and management systems has been set up at CESI Laboratory [14].

Dependability and security modeling
A large body of research exist on the dependability analysis and evaluation of computer based infrastructures, in particular with respect to *accidental threats*, considering two main complementary approaches: i) analytical modeling and ii) measurement-based assessment.

Analytical modeling has been proven to be a cost effective and useful technique to support architectural design decisions, as a model can play the double role of a specification of the design, as well as the basis for a qualitative and quantitative evaluation of the design itself.

Several model-based evaluation techniques are currently available, including combinatorial techniques (reliability block diagrams, fault trees, network graph analysis, Bayesian networks) or state space analysis (Markov chains, stochastic Petri nets and their extensions). State-based models are commonly used, as they are able to capture various functional and stochastic dependencies among components and allow the evaluation of various measures related to dependability and performance. However, the major difficulty with such models is related to the model largeness and state explosion problem. Significant progress has been obtained during the last twenty years, by CRUTIAL partners and elsewhere, for addressing such problem at the model construction and model solution levels. The proposed techniques include the use of hierarchical approaches, composition rules, parametric modelling, state truncation, state aggregation, as well as high-level specification formalisms and automated generation methods, based for example on SPNs (Stochastic Petri Nets) and their extensions, or on semi-formal specification languages like UML diagrams.

As regards *malicious threats*, traditional security evaluation approaches are mainly based on qualitative analysis and evaluation criteria. While these criteria provide useful guidelines during the design, it is widely recognized that they are insufficient to assess the impact of attacks on security during the operation phase. To fill this gap, new approaches have been proposed recently for the quantitative evaluation of security based on probabilistic modelling approaches [15].

Existing results on dependability and security modeling and evaluation have mainly addressed individual infrastructures without tackling the problem of interdependencies between infrastructures. However, to truly understand and analyze the operational characteristics of these infrastructures, it is necessary to model such interdependencies and analyze their impact with respect to the occurrence of critical outages. The need for such modeling and assessment studies has been highlighted in the conclusions of the RamI (R&D challenges for Resilience in Ambient Intelligence") Workshop organized by the European commission in March 2004. Also, it has been pointed out in the recommendations of the roadmap for European R&D on Critical Infrastructure Protection, produced by the ACIP (Analysis & Assessment for Critical Infrastructure Protection) project.

It is only recently that the problem of interdependencies modeling has been addressed by the research community. To the best of our knowledge there are currently no available significant results tackling this problem. The papers published so far on this topic mainly include definitions of the multiple dimensions of interdependencies [3], or discuss potential approaches and preliminary models that are aimed at describing and analyzing the impact of interdependencies (see [16] and references within [17],[18]).

Resilient distributed real-time architectures
There is a reasonable body of research on distributed computing architectures, methodologies and algorithms, both in the fields of dependability and fault tolerance, and in security and information assurance. These are commonly used in a wide spectrum of situations: information infrastructures; commercial web-based sites; embedded systems. Their operation has always been a concern, namely presently, due to the use of COTS, compressed design cycles, openness. Whilst they have taken separate paths until recently, the problems to be solved are of similar nature: keeping systems working correctly, despite the occurrence of mishaps, which we could commonly call faults (accidental or malicious); ensure that, when systems do fail (again, on account of accidental or malicious faults), they do so in a non harmful/catastrophic way. In classical dependability, and mainly in distributed settings, fault tolerance has been the workhorse of the many solutions published over the years. Classical security-related work has on the other hand privileged, with few exceptions, intrusion prevention, or intrusion detection without systematic forms of processing the intrusion symptoms.

A new approach has slowly emerged during the past decade, and gained impressive momentum recently: intrusion tolerance. That is, the notion of -- handling/react, counteract, recover, mask -- a wide set of faults encompassing intentional and malicious faults (we may collectively call them intrusions), which may lead to failure of the system security properties if nothing is done to counter their effect on the system state. In short, instead of trying to prevent every single intrusion, these are allowed, but tolerated: the system has the means to trigger mechanisms that prevent the intrusion from generating a system failure. In fact, the term "intrusion tolerance" was used for the first time more than 20 years ago, and a sequel of that work lead to a specific system developed in the DELTA-4 project [19]. In the following years, a number of isolated works, mainly on protocols, took place that can be put under the IT umbrella [20],[21],[22], but only recently did the area develop significantly, with two main projects OASIS and MAFTIA [23], doing structured work on concepts, mechanisms and architectures. These projects, however, mainly addressed the issue of intrusion tolerance on Internet-oriented infrastructures type of systems. Some progress has been obtained recently by CRUTIAL partners in protocols resilient to malicious faults [24] and on architecting and programming with trusted components [25]. In this project, we intend to make important advances on how to tolerate intrusions on critical utility infrastructures.

**CRUTIAL work plan**
The research that will developed in CRUTIAL address the objective to improve the resilience of the power grid by deploying suited information and communication infrastructures. Two main research directions will be investigated to fulfill this objective. The first research line focuses on the development of a model-based methodology for the dependability and security analysis of the power grid information infrastructures. A second line of research is related to the design of control system architectures and protection mechanisms for electric power systems. Both the (modeling and architectural) project activities are driven by the identification of some *reference control system scenarios* and examples of various types of interdependencies based on the analysis of the existing and ongoing evolutions of the target architectures.

The activities that will be carried out can be summarized as follows:

a) identification and description of reference control system scenarios;
b) development of comprehensive modeling approaches for understanding and mastering the various interdependencies;
c) development of test beds integrating the electric power system and the information infrastructure to support the analysis and validation of control system and interdependencies scenarios;
d) investigation of fault-tolerant architectural configurations and protection mechanisms to enhance the resilience of target infrastructure;
e) provision of qualitative and quantitative support for the identification, analysis and evaluation of the scenarios identified.

The following subsections summarize these activities and the research directions that will be explored.

Identification and Description of Control System Scenarios
A control system scenario defines a reference structure of i) the power grid, ii) the monitoring & control network, with Intelligent Electronic Devices at different levels of the power system (Control Center level, Station level, Bay level, Process level), iii) the management information networks and their functional relationships with the process network, iv) and the different threats that may threaten the operation of the power system services.

Control system scenarios are derived from ongoing evolutions of existing control systems, such as the increasing impact of dispersed electricity generation (renewable energy sources, - e.g. photovoltaic and wind energy- that are connected on different places to the electricity distribution network). If sufficient generation (and storage) facilities are available in a part of the electrical grid, such part can become an energy island (or microgrid) which functions independently from the major grid (e.g. during blackout). Different evolution aspects are taken into account in the scenario identification phase.

A first set of scenarios will be derived from the control of distributed generation systems and microgrids. Such control systems form a so-called Autonomous Electricity Network (AEN), i.e. a group of distributed generators, intelligent loads and storage devices, capable of cooperation and control in a distributed manner, i.e., without central controller, and this based on standard components and public communication networks. External information, such as the instantaneous electricity price from real-time market place, can be incorporated into the control strategies in order to optimize economic control objectives. Hence, this information infrastructure is used for tertiary control, based on communication among the different entities without central coordination. It comes on top of the primary control of voltage and frequency (which is local i.e. without communication) and secondary control (which is centralized and used for e.g. dispatching). As such reliability of these electricity networks is improved, also when sub-networks of the electricity network are separated from the main network. In islanding mode, several issues need to be solved regarding protection and control (e.g. the selectivity of the protection needs to react on different threshold), which requires appropriate communication and control. The ICT infrastructure and several control algorithms are also required when re-synchronizing the microgrid to the main grid after islanding.

A second evolutionary aspect for control scenarios is related to new Power Management Systems which integrate management information networks with process control networks. Security and dependability issues become a major concern in this situation. This also includes the consideration of the configuration capabilities required to substation control systems by the new organization of the power market. These setups address information infrastructures in which involved stakeholders access to the power management system for many different purposes, and some of them require to modify the power network asset through the on-line reconfiguration of the field equipment and their control apparatus.

Interdependencies modelling
The objective is to develop a comprehensive modeling framework aimed at fulfilling the following goals: i) characterize and analyze the interdependencies between the information infrastructure and the controlled power infrastructure, especially the various types of failures that might occur in the presence of accidental and malicious faults; and ii) assess their impact on the resilience of these infrastructures with respect to the occurrence of critical outages and perform sensitivity analyses to support design decisions.

Two complementary analysis and modeling methods will be investigated: a) q*ualitative analysis* methods based on hazard identification and risk modeling techniques, aiming at the identification of failure scenarios, the analysis of their impact and their ranking according to severity and criticality criteria; b) *quantitative evaluation* methods based on stochastic processes (using e.g., Markov chains or stochastic Petri nets), aiming at quantifying the impact of failure and recovery scenarios on the behavior of the infrastructures and the dependability and trustworthiness of the delivered service.

The modeling framework investigated in CRUTIAL is aimed at taking into account multiple dimensions of interdependencies [3]: a) the types of interdependencies (physical, cyber, geographic, logical), b) the coupling and response behavior (loose or tight, inflexible or adaptive), c) the various classes of faults that can occur, and d) the different time scales, and the infrastructure characteristics (organizational, operational, temporal, spatial). Particular focus will be put on the study of the *types of failures* that are characteristic of interdependent critical infrastructures: *Cascading failures, Escalating failures and Common cause failures.*
Although the modeling of such failures has received increasing interest in the last years[9] after the large blackouts of electric power transmission systems in 1996 and 2003, this problem is still open and further developments are needed, in particular with respect to the modeling of escalating failures. The analysis of such types of failures requires the use of models that are able to describe stochastic dependencies and to take into account non-stationary phenomena that result in particular from the combined impact of component failures and performance degradations due to overloads.
Accounting for all such aspects raises a number of challenging issues that need to be addressed and require appropriate methodologies to overcome them. A discussion of some of these challenges and of the research directions that will be explored in the project is given in the following.

A major difficulty lies in the *complexity* of the modeled infrastructures in terms of largeness, multiplicity of interactions and types of interdependencies involved. To address this problem, a number of abstractions and appropriate approaches for composition of models will be necessary. The aim is therefore to produce, from conceptual analyses, generic models that can be refined, instantiated and composed according to hierarchical modeling approaches. Generalization, adaptation and extension of our previous approaches (see [26],[27],[28], [29], [30]) will be necessary to address: i) the specific challenges and characteristics inherent to the combined modeling of information and power distribution infrastructures; ii) the combination of different formalisms to describe the various component of a system and their dependencies); iii) extension of existing formalisms to deal with peculiar features raising from our application context.

Another point is *on-line evaluation* (as opposed to traditional off-line evaluation), which would be desirable in a number of circumstances, in particular to support the efficient selection of adaptive reconfiguration strategies. Of course, although appealing, the online solution shows a number of challenging problems requiring substantial investigations (e.g., dynamic instanciation of models, compositional rules, efficient model solution).

---

[9]  I. Dobson, B.A. Carreras, V. Lynch, D.E. Newman, "Complex Systems analysis of series of blackouts: cascading failure, criticality, and self-organization", *Bulk Power System and Control*, Aug. 2004, Cortina d'Ampezzo, Italy.

An additional relevant issue that needs to be carefully addressed in the modeling concerns the well known stiffness problem, or system parameters that are only partially defined. Such a problem can be alleviated by the use of aggregation and approximation techniques. Also, we need to address the problem of modeling interdependencies in a context characterized by *different operation phases and regimes* with different configurations and behaviors. Such changes might have a significant impact on the parameters describing the occurrence of failures and their propagation. We will investigate the use of stochastic models for multi-phased systems to address this problem [32].

Finally, a difficult issue concerns the evaluation of the impact of *malicious faults*. Traditionally, only accidental faults in software and hardware components have been taken into account in the evaluation of quantitative dependability measures, while the evaluation of security has been mainly based on qualitative evaluation criteria [33]. However, quantitative criteria are widely recognized to be insufficient for analyzing and assessing the impact of malicious attacks and vulnerabilities on the security of systems in operation, or to support the design of intrusion-tolerant systems, and quantitative evaluation based on probabilistic modeling is a promising research direction aimed at filling this gap. Some pioneering work has been achieved by some of the CRUTIAL partners towards this objective [34]. However, to be applicable to the context of interdependent information and power distribution systems, generalization and extensions have to be developed. The ultimate objective that will be pursued during the project on this topic will be the definition of a comprehensive framework for the modeling and evaluation of resilience taking into account malicious faults as well as accidental faults.

Resilient architectures and protection mechanisms
CRUTIAL will study the information infrastructures used in electric power systems, in order to derive common denominators: exposure, threat, vulnerability, unsafety, etc. Thus it will define a reference architecture that takes into account the needs defined by this operation (e.g., that a certain level of exposure is unavoidable). Considering the testbed described above we foresee at least three interconnection realms with different levels of threat, degrees of vulnerability, level of criticality w.r.t. acceptable operational risk: operational SCADA/embedded networks; corporate intranets; Internet/PSTN access.
WP4 will investigate on
- architectural configurations that induce prevention of the more severe interaction faults, and attack and vulnerability combinations. The challenge here, furthering recent research achievements on fault and intrusion tolerant (FIT) architecturing [23], lies exactly with the (unavoidable) entanglement of the information flows of the three above-mentioned realms;
- middleware devices that achieve tolerance of the remaining faults/intrusions (architectural blocks, protocols). We plan to study combinations of recent promising techniques for building protocols that address the different levels of criticality in the foreseen architecture, such as randomization or wormholes;
- sophisticated system monitoring mechanisms that go beyond mere intrusion detection but instead achieve what might be called trustworthiness monitoring, where the general level of resilience of the system is assessed during its operation, both from the perspective of attacks, vulnerabilities, and general faults.

The architectural configurations developed in the project will have to strike a right balance between the prevention and removal of the more severe interaction faults and vulnerability-attack combinations, and the tolerance of the remaining potential faults and intrusions. The challenge here, furthering recent research achievements on fault and intrusion tolerant architecturing, lies exactly with the (unavoidable) entanglement of the information flows of the various above-mentioned infrastructures.

The findings of WP4 investigations will be validated both through proof-of-concept prototypes of the above-mentioned functionality, and real-life testbed experiments, and by building a resilience case in the new architectural scene, and compare it against the starting scenarios studied in the beginning. They will be formally validated also, by several means that include modeling, and experimental validation.

Testbed and experimental evaluation
Two testbed will be developed which integrate the electric power system and the information infrastructure. This will allow the elaboration of new control scenarios in order to better identify them, to allow for the deployment of architectural patterns, and, to complement the modeling to analyze interdependencies and to identify critical failure modes. These testbed platforms will be based on power electronic converters and Intelligent Electronic Devices (IEDs) – such as bay level controllers and power station computers – that are interconnected via off-the-shelf communication protocols (TCP/IP or UDP/IP).

It is clear that in all control scenarios identified, ICT and electric power infrastructures are tightly coupled: a suited architecture needs to be developed and the interdependencies analyzed. The identified scenarios will drive the development of the analysis methodology and of the architectural design. The testbed experiments will instantiate the designed ideas onto innovative infrastructures in which the power process control is connected to the power business management.

The testbed will consist of two platforms; as illustrated below, both platforms are required, because they are complementary to each other.

A *first* testbed consists of several **power electronic converters**, which are interconnected via off-the-shelf communication protocols (TCP/IP). It is based on the Herakles platform [35], developed at K.U.Leuven. Each converter can be used to emulate generators or loads in an energy island. The Herakles platform allows different control ideas (voltage/frequency/current control, power quality control, etc.) to be modeled in a high level programming tool such as Matlab, after which it can be swiftly prototyped on a 4-quadrant power electronic converter, whereby the control algorithms are downloaded on high performance signal processing hardware (DSP + FPGA) which manages the power electronics. This DSP is connected to a PC which allows communication with other intelligent components.

As these Herakles platforms are connected to PC's, they can be interconnected via TCP/IP modules in order to extend the control scope from local towards hierarchical and decentralized control algorithms.

By the project-start, about 3 of these Herakles elements are available, on which mainly local control algorithms are being executed. During the CRUTIAL project, this platform will be further developed to be able to execute different hierarchical and distributed control algorithms. For this, work is required on the hardware and software of the platform, as well as on the middleware modules to interconnect them into a network – as such creating an information infrastructure.

Because of the distributed approach, this platform will be mainly used in the context the so-called 'new' control scenarios. It will be used to evaluate the aspects of interdependencies between the information infrastructure and the electric power system, and to identify the robustness of the control algorithms to disturbances, and to provide feedback to the modeling and architectural parts based on the experiments.

A *second* testbed is based on a platform which is currently under development at CESI labs for supporting the simulation of attack scenarios. The platform consists of an **Ethernet based communication infrastructure interconnecting a power substation control network, a substation supervision network and a control centre network** through protocols based on standard communication stacks (e.g. UDP/IP). The power substation control network is modular and composed by a heterogeneous set of real time targets performing process data elaboration. The software architecture of the station automation application is based on an automation software engineering environment developed at CESI that supports automata-based specification, automatic code generation and distributed execution. The substation supervision network consists of standard elaboration units which provide data base, administration and web services supporting monitoring and maintenance functions, whilst the control center implements remote supervision and control functions, as well as maintenance.

This second set-up is complementary to the first platform because it builds on environments that are used in industrial automation (SCADA-based). It is also well suited to evaluate the effects of fault propagation of e.g. malicious faults, because these SCADA networks can be integrated into corporate control and administrative networks.

**Project Coordinator's Biography**

**Giovanna Dondossola** belongs to the Networks and Infrastructures Department of CESI RICERCA company where she manages European Projects and National activities on dependability and security issues of power network control systems. Among the activities she is currently leading there are interoperability issues in power station control software, fault tolerant distributed architectures in real time industrial automation, cyber security for the electric power system. Two main objectives of the activities on power system electronic security are the development of methodologies for threat analysis and the evaluation of tools for monitoring the communications and for studying attack evolutions and secure architectures.

She is author of about 30 papers published in international conferences proceedings. She is member of the Cigré Working Group D2/B3/C2-01 on Security for Information Systems and Intranets in Electric Power Systems.